
\input phyzzx
\def\cL{{\cal L}}
\def\dplus{=\hskip-5pt \raise 0.7pt\hbox{${}_\vert$} ^{\phantom 7}}
\def\dplusup{=\hskip-5.1pt \raise 5.4pt\hbox{${}_\vert$} ^{\phantom 7}}

\def\dplus{=\hskip-4.8pt \raise 0.7pt\hbox{${}_\vert$} ^{\phantom 7}}

\def\pmb#1{\setbox0=\hbox{#1} \kern-.025em\copy0\kern-\wd0
\kern0.05em\copy0\kern-\wd0 \kern-.025em\raise.0433em\box0}

\def\cM{{\cal M}}
\def\cL{{\cal L}}
\REF\wo{E. Witten and D. Olive, Phys. Lett. {\bf 78B} (1978) 97.}
\REF\bogomol{E.B. Bogomol'nyi, Sov. J. Nucl. Phys. {\bf 24} (1976) 449.}
\REF\Qkinks{E.R.C. Abraham and P.K. Townsend,
Phys. Lett. {\bf 291B} (1992) 85; {\it ibid} {\bf 295B} (1992) 225.}
\REF\wchris{C.M. Hull, G. Papadopoulos and P.K. Townsend, Phys. Lett.
{\bf 316B} (1993) 291.}
\REF\us{G. Papadopoulos and P.K. Townsend, Class. Quantum Grav.
{\bf 3} (1994) 515.}
\REF\ustwo{G.Papadopoulos and P.K. Townsend, Class. Quantum Grav. {\bf 11}
(1994) 2163.}
\REF\AGF{ L. Alvarez-Gaum{\' e} and D.Z. Freedman, Commun. Math. Phys.
{\bf 91} (1983), 87.}
\REF\GH {G.W. Gibbons and S.W. Hawking, Phys. Lett. {\bf 78B} (1978), 430.}
\REF\geometry{S.J. Gates, C.M. Hull and M. Ro{\v c}ek, Nucl. Phys. {\bf B248}
(1984) 157; C.M. Hull, {\it Super Field Theories} ed H. Lee and G. Kunstatter
(New York: Plenum) (1986).}
\REF\gp {P.S. Howe and G. Papadopoulos, Nucl .Phys. {\bf B289} (1987) 264;
Class. Quantum Grav. {\bf 5} (1988) , 1647.}
\REF\hr{N.J. Hitchin, A. Karlhede, U. Lindstr{\"o}m and M. Ro{\v c}ek, Commun.
Math. Phys.
{\bf 108} (1987), 537.}


\Pubnum{ \vbox{ \hbox{R/94/48} } }
\pubtype{}
\date{December, 1994}

\titlepage

\title{Solitons in supersymmetric sigma models with torsion}

\author{G. Papadopoulos}
 \address{DAMTP,\break University of Cambridge, \break Silver Street,\break
Cambridge,
U.K.}
\andauthor{P. K. Townsend}
\address{DAMTP,\break University of Cambridge, \break Silver Street,\break
Cambridge, U.K.}

\abstract { We derive a bound on the energy of the general (p,q)-supersymmetric
two-dimensional massive sigma model with torsion, in terms of the topological
and Noether charges that appear as central charges in its supersymmetry
algebra. The bound is saturated by soliton solutions of first-order
Bogomol'nyi-type equations. This generalizes results obtained previously for
$p=q$ models without torsion. We give examples of massive (1,1) models with
torsion that have a group manifold as a target space. We show that they
generically have multiple vacua and find an explicit soliton solution of an
$SU(2)$ model. We also construct a new class of zero torsion massive (4,4)
models with multiple vacua and soliton solutions. In addition, we compute the
metrics on the one-soliton moduli
spaces for those cases for which soliton solutions are known explicitly, and
discuss their interpretation.}

\endpage

\pagenumber=2



\def\RN{{\cal R}}

\def\half{{1\over2}}

\def\a{\alpha}
\def\b{\beta}

\def\fff{\vrule width0.5pt height5pt depth1pt}
\def\pp{{{ =\hskip-3.75pt{\fff}}\hskip3.75pt }}

\def\cM {{\cal{M}}}



\chapter{Introduction}

For certain supersymmetric field theories, the algebra of supersymmetry
charges can be modified in the presence of a soliton by the appearance
of the soliton's topological charge as a central charge [\wo]. In some
cases the soliton may carry a Noether charge too, and this may also
appear as a central charge in the supersymmetry algebra. Such solitons
are generically referred to as `Q-solitons'. The appearance of central
charges in the supersymmetry algebra implies a bound on the energy of
arbitrary field configurations of the type
$$
E \geq m\sqrt{ \sum Q^2} \ ,
\eqn\bogoone
$$
for suitably normalized charges, $Q$, topological or Noether, where $m$ is a
mass parameter. Bounds of this type were originally derived by Bogomol'nyi
[\bogomol] for various bosonic field theories but without mention of
supersymmetry. Bogomol'nyi's arguments have a natural interpretation in terms
of
supersymmetry, however, and it is only in this context that the bound can be
expected to survive quantum corrections. The bound is saturated by solutions to
first-order equations which imply, but are much simpler than, the second-order
field equations. The solutions of these first-order, Bogomol'nyi, equations are
precisely the soliton, or Q-soliton, solutions. An example of a Q-soliton in a
four-dimensional supersymmetric field theory that can be found in this way is
the BPS dyon of $N=2$ super Yang-Mills theory.

As is well-known, two-dimensional non-linear sigma-models are analogous in many
respects to four-dimensional gauge theories. The two-dimensional analogues of
the magnetic monopole and dyon are the kink and $Q$-kink [\Qkinks] solutions of
sigma models with a scalar potential, which we shall call `massive'
sigma-models
since the presence of the scalar potential introduces a mass parameter.
Omitting fermions, the action takes the form
$$
 I=\int dxdt\, \Big[ (g+b)_{ij} {\partial_\pp}\phi^i
\partial_= \phi^j  - V(\phi)\big] \ ,
\eqn\aaone
$$
where $\phi$ is a map from the two-dimensional Minkowski space-time,
with light-cone co-ordinates $(x^\pp,x^=)$, into the target manifold $\cM$
with metric $g$. The two-form $b$ is a locally-defined potential
for a globally-defined `torsion' three-form $H$ with components
$H_{ijk}={3\over2}\partial_{[i}b_{jk]}$, and $V$ is the scalar potential.

The charges carried by the kink or $Q$-kink solutions of the
(p,q)-supersymmetric sigma-model field equations appear as central charges in
the (p,q)-supersymmetry algebra; note that this is true not only of the
topological charges but also of the Noether charges. Lorentz-invariant central
charges are possible only if neither $p$ nor $q$ vanishes, so the sigma-models
of relevance here have at least (1,1) supersymmetry. All such models have a
scalar potential of the form [\wchris]
$$
V= {1\over4}m^2 g^{ij}(u-X)_i(u-X)_j\ ,
\eqn\pot
$$
where $m$ is the mass parameter, $X$ is a (possibly vanishing) Killing vector
field on $\cM$ and $u$ is a one-form on $\cM$ whose exterior derivative $du$ is
determined by $X$ and $H$ via the formula
$$
 X^k H_{kij}=\partial_{[i} u_{j]}\ .
\eqn\defu
$$
In particular, $du=0$ if either $H$ or $X$ vanishes in which case $u=da$
for some locally-defined scalar $a$ on $\cM$, which can be identified as
the superpotential of the (1,1)-superspace formulation. If $X$ is
non-vanishing then $H$ must be invariant under the symmetry generated by
$X$, i.e. $\cL _X H=0$. The conditions $\cL _X H=0$ and $dH=0$ imply only that
$X\cdot u=const.$ but consideration of the closure of the Poisson bracket
algebra of charges for Minkowski space-time requires that
$$
X\cdot u=0\ .
\eqn\defua
$$
These requirements imply that $u$, and hence the scalar potential $V$, is
invariant under the symmetry generated by $X$.

A limitation of all models for which the soliton structure has been
investigated to date is that they are non-chiral. That is, the number of left
and right-handed supersymmetries is equal, i.e. $p=q$, and there is no
parity-violating Wess-Zumino, or `torsion', term in the action. Recently, we
have determined the conditions imposed by {\it on-shell} (p,q)
supersymmetry\foot{These conditions were found for both off-shell and on-shell
supersymmetry. The conditions for off-shell supersymmetry are more restrictive
than those of on-shell supersymmetry.} on the form of the potential $V$ in the
general massive sigma model with torsion [\us,\ustwo], generalizing the
off-shell (p,0) and (1,1) results of [\wchris] and the earlier results
for (p,p) models without torsion [\AGF]. One purpose of this paper is to
present some examples of massive sigma models, with and without torsion, for
which these conditions are satisfied and which are therefore
(p,q)-supersymmetric. In particular, we present massive extensions of the
(1,1)-supersymmetric WZW models and a class of (4,4) models with hyper-K\"ahler
(4k)-metrics admitting k commuting tri-holomorphic Killing vector fields. The
latter examples generalize the k=1 models based on the Gibbons-Hawking
4-metrics [\GH].

Our main interest in these models in this paper is the possibility of soliton
solutions interpolating between isolated zeros of the potential $V$, an example
of which is the Q-kink solution found in [\Qkinks] for a (4,4) supersymmetric
sigma model. Using the results of [\ustwo] for the Poisson bracket algebra of
supersymmetry charges, reviewed in section 2, we derive in section 3 a
Bogomol'nyi bound on the energy of a general field configuration, and we show
that the bound is saturated by solutions of first order equations of
Bogomol'nyi type. The finite energy solutions of these equations are the
soliton solutions of these models. In subsequent sections we investigate in
detail various
special cases and find a new explicit soliton solution in a massive (1,1)
models with torsion and the $SU(2)$ group as its target space. We also compute
the metric on the one-soliton moduli space in both this example and the Q-kink
example of [\Qkinks], and hence the one-soliton low-energy effective
Lagrangian. In the Q-kink example this Lagrangian has a Kaluza-Klein
interpretation.


\chapter{ Massive supersymmetric sigma-models}

We begin with a summary of those results of refs. [\wchris,\us,\ustwo] that
will be relevant to the derivation of the Bogomol'nyi bound in the following
section. The component  action of the general (1,1)-supersymmetric model is a
functional of the sigma model fields $\phi^i$ and the chiral fermions
$\lambda_+$ and $\psi_-$, and takes the form
$$
\eqalign{
I =\int\! d^2 x\big\{ &\partial_\dplus\phi^i\partial_=\phi^j (g_{ij}+b_{ij})
+ ig_{ij}\lambda_+^i\nabla_=^{(+)}\lambda_+^j -
ig_{ij}\psi_-^i\nabla_\pp^{(-)}\psi_-^j\cr
& -{1\over2}\psi_-^k\psi_-^l\lambda_+^i\lambda_+^j R^{(-)}_{ijkl}
 +m\nabla^{(-)}_i (u-X)_j \lambda_+^i\psi_-^j  -V(\phi)
\big\}\ ,}
\eqn\ssaction
$$
where $V$ is given in eqn. \pot, and $\nabla^{(\pm)}$ are the covariant
derivatives with connections
$$
\Gamma^{(\pm)}_{ij}{}^k = \big\{\matrix{ k\cr ij\cr}\big\}
\pm  H_{ij}{}^k\ ;
\eqn\torsion
$$
i.e. $H_{ijk}$ is the torsion of the connection of $\nabla^{(+)}$.  We refer to
[\us,\ustwo ] for details of the conventions.

All $(p,q)$-supersymmetric sigma models with $p,q\geq 1$ are special cases of
the (1,1)-supersymmetric model. The additional supersymmetries simply impose
further restrictions on the sigma model couplings and the geometry of $\cM$. In
the massless case these restrictions are long-established [\geometry,\gp]. For
example, an additional $p-1$ left-handed supersymmetries requires the existence
of $p-1$ complex structures $I_r$ on $\cM$ that are covariantly constant with
respect to the connection $\Gamma^{(+)}$, and that the metric $g$ of $\cM$ be
hermitian with respect to them. In the case that $p=4$, closure of the algebra
of supersymmetry transformations requires in addition that the complex
structures $I_r$ ($r=1,2,3$) obey the algebra of imaginary unit quaternions.
Similarly, an additional $q-1$ right-handed supersymmetries requires the
existence of ($q-1$)-complex structures $J_s$ on $\cM$, but in this case the
complex structures $J_s$ are covariantly constant with respect to the
connection $\Gamma^{(-)}$. The metric $g$ must be Hermitian with respect to all
complex structures.

Apart from the complex structures $I_r$ and $J_s$, the massive
(p,q)-supersymmetric sigma models also depend on a set of Killing-vector/one
-form pairs $\{Z_{I'I}, u^{I'I}\}$. The Killing vector fields $Z_{I'I}$
leave the torsion $H$ and the complex structures $I_r$ and $J_s$ invariant and
generate new symmetries of the action, and they also have the same relationship
with the one-forms $u^{I'I}$ as $X$ has to $u$, \rm{i.e}
$$
Z_{I'I}\cdot H=du^{I'I}.
\eqn\bone
$$
For consistency with [\ustwo] we adopt the notation
$$
\eqalign{
Z_{00} &=X \qquad Z_{0r} =Z_r \qquad Z_{s0} =Y_s \qquad Z_{sr} =Z_{sr}
\cr
  u^{00}&=u\qquad    u^{0r}=v_r  \qquad  u^{s0}=w_s  \qquad u^{sr}=v_{sr}\ .}
\eqn\defaa
$$
The most important restrictions imposed by (p,q) supersymmetry can now be
summarized by the following set of relations:
$$
\eqalign{
(Z_r + v_r)_i + I_r{}^k{}_i (X+u)_k &= 0
\cr
(Y_s-w_s)_i +  J_s{}^k{}_i(X-u)_k&=0
\cr
(Z_{sr} + v_{sr})_i + I_r{}^k{}_i (Y_s + w_s)_k &=0
\cr
(Z_{sr} - v_{sr})_i + J_s{}^k{}_i (Z_r - v_r)_k &=0 }
\eqn\btwo
$$
and
$$
Z_{I'I}\cdot u^{J'J}+Z_{J'J}\cdot u^{I'I}=0\ ,
\eqn\bthree
$$
which includes \defua\ as a special case. For a discussion of the details see
[\ustwo].

The supersymmetry charges of these models consist of the (1,0) and (when
applicable) extended (p,0) charges, $S_+$ and $S_+^r$, and the (0,1) and
extended (0,q) charges, $S_-$ and $S_-^s$. The other, bosonic, charges
appearing in the supersymmetry algebra are the energy, $E$, the momentum, $P$,
and the central charges $Q_{I'I}$. These charges have the following
expressions\foot{This corrects the omission in [\ustwo] of a torsion term in
$S_+$.} in terms of the sigma-model fields:
$$
\eqalign{
E &={1\over 2} \int\! dx [g_{ij} \partial_t\phi^i
\partial_t\phi^j + g_{ij} \partial_x\phi^i
\partial_x\phi^j + V(\phi) + \ {\rm fermions}\ ]\cr
P&=\int\! dx [g_{ij} \partial_t\phi^i \partial_x\phi^j+ {\rm fermions}\ ]\cr
Q_{I'I}&=\int\! dx [Z^{I'I}{}_i \partial_t\phi^i
+ u^{I'I}{}_i \partial_x\phi^i + {\rm fermions}]\cr
S_+&=\int\! dx [g_{ij}
\partial_\pp\phi^i \lambda^j_+
-{1\over3}H_{ijk}\lambda_+^i\lambda_+^j\lambda_+^k
-{i\over 2} m (u-X)_i \psi^i_-]\cr
S_-&=\int\! dx [ i g_{ij} \partial_=\phi^i \psi^j_-
 + {1\over 3} \psi^i_- \psi^j_- \psi^k_-
H_{ijk} + {m\over 2} (X+u)_i \lambda_+^i]\cr
S^r_+&=\int\! dx [ I_{r\; ij} (\partial_\pp \phi^i \lambda_+^j-
 i H^i{}_{kl} \lambda^j_+
\lambda^k_+ \lambda^l_+) - i{m\over 2} (v_r-Z_r)_i \psi^i_-]\cr
S^s_-&=\int\! dx [i J_{s\; ij} \partial_=\phi^i \psi^j_-
+ {1\over 3} H_{mnl} J_s{}^m{}_i J_s{}^n{}_j J_s{}^l{}_k \psi^i_-  \psi^j_-
\psi^k_- +{m\over 2} (Y_s+w_s)_i \lambda_+^i]\ .}
\eqn\abfour
$$
The fermion contributions to the bosonic charges will not be needed for what
follows so we have omitted them.

Rewriting the above charges in terms of the sigma model fields and their
conjugate momenta, and using the canonical Poisson bracket relations, one finds
the following Poisson Bracket algebra of the charges:
$$
\eqalign{
\{S_+,S_+\}&= 2 (E+P), \quad \{S^r_+,S^s_+\}= 2 \delta_{rs} (E+P),
 \quad \{S_+,S^r_+\}=0 ,
\cr
\{S_-,S_-\}&=2(E-P),
\quad \{S^r_-,S^s_-\}= 2 \delta_{rs} (E-P), \quad \{S_-,S^s_-\}=0 ,
\cr
\{S_+,S_-\}&=m Q_{oo}, \quad \{S_+,S_-^s\}= m Q_{so},
\cr
\{S_-,S_+^r\}&= m Q_{or}, \quad \{S_+^r,S_-^s\}= m Q_{sr}\ .}
\eqn\absix
$$
In addition, the Poisson brackets of the $Q_{I'I}$ amongst themselves, and with
all supersymmetry charges, vanish, i.e. the $Q_{I'I}$ are central charges of
the
supersymmetry algebra.


\chapter{Bogomol'yni Bounds}

Using the algebra of Poisson brackets just reviewed, the energy $E$ can be
written as
$$
\eqalign{
E=&\{\a_+ S_+-\a_- S_- +\b_{+{}r} S^r_+-\b_{-{}s} S^s_-\ ,\ \a_+ S_+-\a_- S_-
+\b_{+{}r} S^r_+-\b_{-{}s} S^s_-\}+
\cr
&m \Big(\a_+ \a_- Q_{oo} + \a_+ \b_-^s Q_{so} +\a_- \b_+^r Q_{or} + \b^r_+
\b^s_- Q_{{sr}}\Big)}
\eqn\energy
$$
where  $\{\a_-,\a_+,\b_-^s,\b_+^r\}$ are real constants restricted by
$$
\a_+^2+ \sum_{r=1}^{p-1} \big(\b_+^r\big)^2 =\half \qquad
\a_-^2+\sum_{s=1}^{q-1}\Big(\b_-^s\Big)^2 =\half\ .
\eqn\restrict
$$
Using in \energy\ the expressions given in \abfour\ for the supersymmetry
charges, and evaluating the resulting expression by using the canonical
Poisson brackets of the fields with their conjugate momenta, we find (omitting
fermion terms) that
$$
\eqalign{
{E}=\int\! dx \Big[ g_{ij} \Big(\partial_\pp\phi^i &
-  m (\a_+ \delta^i_k+\b^r_+
I_r{}^i{}_k) (\a_- (X+u)^k + \b_-^s (Y_s+w_s)^k) \Big)
\cr
\Big(\partial_\pp\phi^j &-  m (\a_+
\delta^j_l+\b^r_+ I_r{}^j{}_l) (\a_- (X+u)^l + \b_-^s (Y_s+w_s)^l) \Big) +
\cr
g_{ij} \Big(\partial_=\phi^i & +  m (\a_- \delta^i_k+\b^s_-
J_s{}^i{}_k) (\a_+ (u-X)^k + \b_+^r (v_r-Z_r)^k) \Big)
\cr
 \Big(\partial_=\phi^j &+  m (\a_-
\delta^j_l+\b^s_- J_s{}^j{}_l) (\a_+ (u-X)^l + \b_+^r (v_r-Z_r)^l)\Big)\Big]+
\cr
m \Big(\a_+& \a_- Q_{oo} + \a_+ \b_-^s Q_{so} +\a_- \b_+^r Q_{or} + \b^r_+
\b^s_- Q_{{sr}}\Big)}
\eqn\afive
$$
where
$(\partial_\pp,\partial_=)  = (\partial_t+\partial_x ,\partial_t-\partial_x)$.

An immediate consequence of \afive\ is that
$$
{E} \geq {m\over2}\ \xi\cdot {\cal Q}   \eqn\nice
$$
where $\xi= 2(\a_-\a_+,\a_+\b_-^s,\a_-\b_+^r,\b_-^s\b_+^r)$ is a ($pq$)-vector
of unit length (as result of \restrict ) and ${\cal Q}=( Q_{oo}, Q_{so},
Q_{or},
Q_{{sr}})$. By choosing $\xi$ parallel to ${\cal Q}$ we derive the
energy bound
$$
{{E}}\geq \half |m|
{\sqrt{|Q_{oo}|^2+\sum_s|Q_{so}|^2+\sum_r|Q_{or}|^2+\sum_{s,r}|Q_{{sr}}|^2}}
\eqn\asix
$$
which is saturated by solutions of the first-order, `Bogomol'nyi' equations
$$
\partial_t\phi^i=  {\cal {X}}^i, \qquad \partial_x\phi^i=  {\cal Y}^i \ ,
\eqn\etwo
$$
where ${\cal {X}}$ and ${\cal {Y}}$ are the vector fields
$$
\eqalign{
{\cal {X}}^i&\equiv m (\a_+ \a_- X^i + \a_+ \b_-^s Y_s^i +\a_- \b_+^r
Z_r^i + \b_+^r \b_-^s Z_{sr}^i)
\cr
{\cal Y}^i&\equiv m(\a_+ \a_- u^i + \a_+ \b_-^s w_s^i +\a_- \b_+^r
v_r^i + \b_+^r \b_-^s v_{sr}^i)\cr
&1\le r\le p \qquad 1\le s\le q  \ .}
\eqn\eone
$$

It can be shown, using the eqns. \bthree,  that the vector fields ${\cal {X}}$
and ${\cal Y}$ are orthogonal and commute under Lie brackets. Therefore the
flows of the vector fields ${\cal {X}}$ and ${\cal Y}$ commute and are
orthogonal. The most general solution of \etwo\ is given by
$$
\phi(x,t)= {{\cal{F}}^{\cal{X}}}_t \big( {{\cal{F}}^{\cal Y}}_x (p)\big)
={{\cal{F}}^{\cal{Y}}}_x \big( {{\cal{F}}^{\cal{X}}}_t (p)\big), \qquad
p\in {\cal{M}}, \eqn\ethree
$$
where ${{\cal{F}}^{\cal{X}}}, {{\cal{F}}^{\cal Y}}$ are the flows of the
vector fields ${\cal{X}}$ and ${\cal{Y}}$ respectively such that
${\cal{F}}^{\cal{X}}_o(p)=p$ and  ${\cal{F}}^{\cal Y}_o(p)= p$. Therefore the
different solutions of \etwo\ are parameterised by the points $p$ of the sigma
model target manifold ${\cal {M}}$.  Note that  {\it both} vector fields
${\cal{X}}$ and ${\cal{Y}}$ vanish at the points $\{p_0\}$ for which the
potential $V$ vanishes, i.e. at the classical vacua; in this case
$\phi(x,t)=p_o$. Other finite energy solutions include those which interpolate
between these classical vacua, but those solutions which merely interpolate
between two zeros of ${\cal Y}$ also have finite energy. In some cases there
are `cosmological' solutions for which $\phi$ depends only on time.

A metric can be defined on the space of solutions, $\phi$, of the Bogomol'nyi
equations as
$$
ds^2={1\over2}\int\! dx\, g_{ij}(\phi^i) \ d\phi^i\ d\phi^j,
\eqn\efour
$$
where $d\phi$ is a differential of $\phi$ with respect to the parameters of the
solutions.
Using that ${\cal {X}}$ is a Killing vector field, so its flow leaves invariant
the sigma
model metric $g$, the metric \efour\ can be rewritten as
$$
ds^2={1\over2}\int\! dx ({{\cal{F}}_x^{\cal Y}}^*g)_{ij}(p) dy^i(p) dy^j(p)
\eqn\efive
$$
where $y^i$ is a set of coordinates on $\cM$ at $\phi(0,0)=p$.  The metric
$ds^2$ is not well-defined for all solutions of the Bogomol'nyi equations
because the expression in equation \efive\ can become infinite, but it is
well-defined for soliton solutions, as we shall confirm.


\chapter{(1,1) Models: the massive supersymmetric WZW model.}

 For (1,1) models eqn. \eone\ becomes
$$
 \partial_x\phi^i\pm {m\over2}  u^i=0 \qquad
 \partial_t\phi^i\pm {m\over2}  X^i=0\ .
 \eqn\sone
$$
As mentioned in the previous section, these equations describe the flows of two
orthogonal commuting vector fields $X$ and $u$, so the two equations are
simultaneously integrable. The solutions of these equations are  $\phi(x,t)=
{{\cal F}^X}_t {{\cal F}^u}_x (p)$ ($p\in {\cal M}$), and those with finite
energy must interpolate between zeros of $u$. Those solutions, parametrised by
the points $p\in {\cal M}$, for which the vector field $X$ vanishes, are
static.

When the torsion $H$ vanishes everywhere on $\cM$ (but not, necessarily, $X$)
the first equation in \bone\ becomes $\partial_x\phi^i\pm {m\over2}
g^{ij}\partial_ja=0$ where $u=da$. The Noether charge
$$
Q_X=\int\! dx (X_i\partial_t\phi^i+u_i\partial_x\phi^i)
\eqn\stwo
$$
associated to $X$ is now naturally expressed as the sum of a new Noether charge
$$
Q'_X=\int\! dx \, X_i\partial_t\phi^i
\eqn\sthree
$$
and a topological charge
$$
Q_a=\int\! dx\, \partial_x\phi^i \partial_ia\ .
\eqn\sfour
$$
For soliton solutions the value of the latter is $a(p_1)-a(p_2)$ where $p_1$
and $p_2$ are the two critical points of $a$ between which the solution
interpolates.

When the Killing vector field $X$ vanishes everywhere on ${\cal M}$ (but not,
necessarily, $H$) $u=da$ for some (locally) defined function $a$. The finite
energy solutions then interpolate between critical points of $a$,
which may be identified as the superpotential of the (1,1)-superspace
formulation of the model. The charge $Q_X$ is topological and its value is
given by $Q_X =a(p_1)-a(p_2)\equiv Q_a$, where $p_1$ and $p_2$ are the critical
points of $a$ that are interpolated by the solution. In principle, there might
be more than one solution interpolating between a given two critical points, or
none. When $a$ is a Morse function we can define as `elementary' solitons those
that interpolate between critical points with Morse indices that differ by one.

Examples of models with both $X$ and $H$ non-vanishing may be found by
consideration of sigma models with a group manifold as their target space.  Let
$K$ be a group manifold with Lie algebra $\cL(K)$ and let $L^A$ and $R^A$,
defined by
$$
k^{-1}dk=L^A t_A\qquad dk\, k^{-1}=R^A t_A, \qquad k\in K\ ,
\eqn\seone
$$
be the left and right invariant frames, respectively, where $\{t_A\}$ is a
basis in $\cL(K)$, $[t_A,t_B]=f_{AB}{}^Ct_C$, and $f_{AB}{}^C$ are the
structure constants.  Note the Maurer-Cartan equations satisfied by the frames:
$$
dL^A=-{1\over2} f_{BC}{}^AL^BL^C\qquad  dR^A={1\over2} f_{BC}{}^AR^BR^C\ .
\eqn\setwo
$$
We choose the sigma model metric $g$ and torsion $H$ to be the bi-invariant
tensors
$$
\eqalign{
g_{ij}&=h_{AB} L^A_i L^B_j=h_{AB} R^A_i R^B_j
\cr
H_{ijk}&=-{\lambda\over 2} f_{ABC}L^A_i L^B_j L^C_k=-{\lambda\over 2}
f_{ABC}R^A_i R^B_j
R^C_k\ ,}
\eqn\sethree
$$
where $h_{AB}$ is a non-degenerate invariant quadratic form on $\cL(K)$,
$f_{ABC}=f_{AB}{}^Dh_{DC}$ and $\lambda$ is a real number. We choose the
Killing vector field $X$ such that
$$
X^i=\kappa^A(L^i_A-R_A^i)\ ,
\eqn\sefour
$$
where $\kappa^A$ are the components of $\kappa=\kappa^At_A $ in $\cL(K)$.
This vector field is generated by the adjoint action,
$$
k\rightarrow e^{y\kappa} k\, e^{-y\kappa}\ ,
\eqn\sefive
$$
of the one-parameter subgroup of $K$ parametrized by $y$. The one-form $u$
defined by \defu\ is then given by
$$
u=\lambda \kappa_A(L^A+R^A)+ da
\eqn\sesix
$$
where $a$ is a function on $K$ (possibly defined only locally) and $\kappa_A =
h_{AB}\kappa^B$. The orthogonality relation $X\cdot u=0$ (eqn \bthree) now
implies that
$$
X^i\partial_ia=0\ .
\eqn\seseven
$$
Since it is clearly possible to find invariant functions $a$ on $K$, for
example the trace of any group element, we have now established the existence
of (1,1) models with torsion having a scalar potential
$$
V= {m^2\over4} \Big( \kappa_A\big[ (\lambda -1) L^A + (\lambda+1)R^A\big] +
da\Big)^2\ .
\eqn\seeight
$$
We note for future reference that the fermion mass matrix $M_{ij} =
m\nabla^{(-)}_i(u-X)_j$ is given by
$$
M_{ij} = m \nabla^{(-)}_i\Big(\kappa_A\big[ (\lambda -1) L_j^A +
(\lambda+1) R_j^A\big] + \partial_j a\Big) \ .
\eqn\seeighta
$$
For $|\lambda|=1$ the connection $\Gamma^{(\pm)}$ is the paralellizing
connection with respect to the left (+) or right (-) invariant frames, so this
model is then a massive extension of the massless WZW model for the group $K$.

Our main interest here is in models for which the scalar potential $V$ has
multiple supersymmetric vacua; for the massive WZW models just constructed
these are the points $k$ of $K$ for which $V=0$. When $\lambda=1$ they are
given by
$$
2\kappa_AR^A + da =0\ .
\eqn\senine
$$
When $a$ is chosen to be
$$
a={1\over n}\tr k^n\ ,
\eqn\asenine
$$
where $n$ is a positive integer, \senine\ becomes
$$
2\kappa_A + \tr k^n t_A =0\ .
\eqn\seten
$$
Since $\kappa_A$ is part of the definition of the model, this equation is to be
solved for the group element $k$. We shall now investigate this equation for
the group $K=SU(2)$. Let
$$
k^n = \pmatrix{A&B\cr -{\bar B}&\bar{A}};\qquad A\bar{A}+B\bar{B}=1\ .
\eqn\seeleven
$$
Then, for a standard choice of basis for $\cL\big(SU(2)\big)$, \seten\ becomes
$$
\eqalign{
2\kappa_1 + i(A-\bar{A}) =0\cr
2\kappa_2 +i( B-\bar{B}) =0\cr
2\kappa_3 + (B+\bar{B})=0}
\eqn\setwelve
$$
for which the solution is
$$
\eqalign{
k^n &= \mp \sqrt{1-|\kappa|^2}\, {\bf 1} + \kappa^A t_A\cr
    &=i\pmatrix{ (\pm
i\sqrt{1-|\kappa|^2}+\kappa_1 )& (\kappa_2 +i \kappa_3)\cr (\kappa_2
-i\kappa_3)&  (\pm
i\sqrt{1-|\kappa|^2} -\kappa_1)}\ .}
\eqn\sethirteen
$$
where
$$
|\kappa|^2=\kappa_1^2+\kappa_2^2+\kappa_3^2\ .
\eqn\setwelvea
$$
There are three cases to consider: Case (i) $|\kappa|>1$: the matrix $k^n$ is
not unitary and hence there are no supersymmetric vacua.  Thus, supersymmetry
is spontaneously broken. This shows, incidentally, that the Witten index
vanishes. This was to be expected because the Witten index of a
(1,1)-supersymmetric sigma model is the Euler number of the target manifold
which vanishes for a group manifold.  Case (ii) $|\kappa|=1$: the matrix $k^n$
is in $SU(2)$ and (by diagonalization) one can prove that there are exactly $n$
supersymmetric vacua, counting multiplicities. In particular, there is
precisely one supersymmetric vacuum state for $n=1$ and so one state of zero
energy. This result is
consistent with the vanishing of the Witten index because all these vacua are
degenerate in the sense that the fermion mass matrix (in the vacuum) is not
invertible, so some fermions remain massless. To see this note first that when
$\lambda=1$ \seeighta\ simplifies to
$$
M_{ij} = m \nabla^{(-)}_i\partial_j a\ .
\eqn\setwelveb
$$
For the choice of $a$ leading to the vacua $k$ given by \sethirteen\ this
becomes
$$
M_{ij} = m R_i^A R_j^B \tr \big( t_A t_B k^n\big)
\eqn\setwelvec
$$
and given the form of $k^n$ in \sethirteen\ it is easily verified that $\det M$
vanishes if and only if $|\kappa|=1$. Case (iii) $|\kappa|<1$: the matrix $k^n$
is again in $SU(2)$ and (again by diagonalization) one can show that there are
exactly $2n$ points $k$ in $K$, counting multiplicity, for which $V=0$, and
hence $2n$ (non-degenerate) supersymmetric vacua. In particular there are
precisely two supersymmetric vacua for $n=1$. We remark that this case is
similar to what happens for $\lambda=0$, i.e. vanishing torsion, in that there
too one finds $2n$ supersymmetric vacua for $a={1\over n}\tr k^n$.

We now turn to the Bogomol'nyi equations \sone\ for the massive WZW models
considered above. These are
$$
\eqalign{
\partial_t k (x,t) &\pm {m\over2} (\kappa k -k\kappa) =0\cr
\partial_x k(x,t) &\pm {m\over2}\Big[\kappa k + k\kappa + k\sum_A t_A\tr (t_A
k^n)\Big]=0
\ .}
\eqn\sefourteen
$$
The first of these equations can be solved by setting
$$
k(x,t)=e^{\mp {m\over2}\kappa t} k(x) e^{\pm {m\over2}\kappa t} \ .
\eqn\sefourteena
$$
The time dependence now cancels from the second equation which reduces to the
ordinary differential equation
$$
{dk(x)\over dx} \pm {m\over2}\Big[\kappa k(x) + k(x)\kappa + k(x)\sum_A t_A\tr
(t_A
k^n(x))\Big]=0
\eqn\sefourteenb
$$
for $k(x)$, the finite energy solutions of which are the solitons.

We shall give an example of a soliton solution for $K=SU(2)$ with $a=\tr k$,
i.e. n=1, and
$$
\kappa =(0, \kappa, 0); \qquad |\kappa|<1\ .
\eqn\sefifteen
$$
{}From \sethirteen\ one sees that the supersymmetric vacua in this case are
$$
k=i\pmatrix{ \pm i\sqrt{1-\kappa^2}& \kappa \cr \kappa
& \pm i\sqrt{1-\kappa^2}}\ ,
\eqn\sefifteena
$$
To find soliton solutions we parametrise $k(x)$ as in \seeleven, i.e.
$$
k(x) = \pmatrix{A(x)&B(x)\cr -{\bar B}(x)&\bar{A}(x)};\qquad
A\bar{A}+B\bar{B}=1\ .
\eqn\sefifteenb
$$
Equation \sefourteenb\ can now be rewritten as
$$
\eqalign{
-{d\over dx}(A-\bar{A})&\pm {m\over2} (A+\bar{A}) (A-\bar{A})=0
\cr
-{d\over dx}(B+\bar{B})&\pm {m\over2} (A+\bar{A}) (B+\bar{B})=0
\cr
{d\over dx}(A+\bar{A})&\pm{m\over2}\big[2i \kappa
(B-\bar{B})+4-(A+\bar{A})^2\big]=0
\cr
{d\over dx}(B-\bar{B})&\pm{m\over2}\big[2i \kappa
(A+\bar{A})-(A+\bar{A})(B-\bar{B})\big]=0\ .}
\eqn\sesixteen
$$
The solutions of these equations that interpolate between the two vacua are
$$
\eqalign{
A&= \sqrt{1-\kappa^2}\big\{- \tanh\big(\pm m \sqrt{1-\kappa^2}\,
(x-x_0)\big)+i\cos(\theta)\, [\cosh \big(m \sqrt{1-\kappa^2}\,
(x-x_0)\big)]^{-1}\big\}
\cr
B&=i \kappa +\sqrt{1-\kappa^2}\, \sin(\theta)\, [\cosh \big(m \sqrt{1-\kappa^2}
(x-x_0)\big)]^{-1}\ ,}
\eqn\seseventeen
$$
where $(x_0,\theta)$, with ranges $-\infty<x_0<+\infty$, $0\leq \theta<2 \pi$,
parameterize the solutions. That these solutions interpolate between the vacua
\sefifteena\ can be verified by analysis of their asymptotic  behaviour as
$x\rightarrow \pm\infty$. To obtain the full solution $k(x,t)$ of the
Bogomol'yni equations, we simply substitute $k(x)$ of \seseventeen\ into
\sefourteena\ with the vector $\kappa$ given by \sefifteen. A computation of
the charge $Q_X$ for these solitons reveals that
$$
Q_X =\mp  4(1-\kappa^2)^{3\over2}
\eqn\seeighteen
$$
which means that the energy is
$$
E=  2m(1-\kappa^2)^{3\over2}
\eqn\senineteen
$$
as can be verified explicitly. Because the two supersymmetric vacua degenerate
at $|\kappa|=1$ the soliton solution reduces to a constant at these values of
$\kappa$, and the energy must vanish, as indeed it does.

The moduli space of the soliton solutions found above is topologically a
cylinder because the solutions depend on the angle $\theta$ in addition to
their position, $x_0$, in space. The 2-metric $ds^2$ of \efour\ on
the moduli space is
$$
ds^2={1\over m} \sqrt{1-\kappa^2}\, d\theta^2+m (1-\kappa^2)^{3/2} dx_0^2\ .
\eqn\seeighteen
$$
Defining the new variable $y$ by
$$
\theta = m\sqrt{1-\kappa^2}\, y
\eqn\senineteen
$$
we observe that
$$
\Big({ds\over dt}\Big)^2 = {1\over2}E (\dot x_0^2 + \dot y^2)\ ,
\eqn\setwenty
$$
which can be interpreted as the bosonic part of the effective low-energy
worldline Lagrangian for the soliton (with mass equal to $E$, as expected). The
full effective Lagrangian will include the coefficients of the fermionic zero
modes in the soliton background as worldline fermions, and will have an $N=1$
worldline supersymmetry.


\chapter {(2,2) Models}

For (2,2) models the Bogomol'nyi equations \eone\ become
$$
\eqalign{ \partial_t\phi^i-& m
(\a_+\a_- X^i+\a_+\b_- Y^i+\a_-\b_+ Z^i+\b_+\b_- T^i)=0
\cr
\partial_x\phi^i-& m
(\a_+\a_- u^i +\b_+\b_-
n^i+
\a_+
\b_- w^i + \a_-\b_+
v^i)=0\ ,}
\eqn\sfive
$$
where $n=v_{11}$.

In the case that all vector fields $X,T,Y,Z$ vanish, we set $u=da, v=db, w=dc$
and $n=de$. To find non-trivial solutions of \btwo\ in this case we shall
suppose that $I$ and $J$ commute, in which case $b,c,e$ can be expressed in
terms of $a$ as follows:
$$
\eqalign{
\partial_ib&=-I^j{}_i\partial_ja, \qquad \partial_ic=-J^j{}_i\partial_ja
\cr
\partial_ie&=(IJ)^j{}_i\partial_ja\ .}
\eqn\ssix
$$
The Bogomol'nyi equations then become
$$
\eqalign{
\partial_t\phi^i&=0
\cr
\partial_x\phi^i&= m\big[\a_+\a_- g^{ij} +\b_+\b_- (IJ)^{ij} + \a_+ \b_-
J^{ij} +\a_-\b_+ I^{ij}\big]\partial_j a\ .}
\eqn\sseven
$$
Since $I$ and $J$ commute their product $\Pi \equiv IJ$ is a product structure
on the sigma model manifold $\cM$. We can choose co-ordinates $\{x^i\}=\{y^a,
z^p\}$ on $\cM$ such that $\{\Pi^i_j\}=\{\delta^a_b, -\delta^p_q\}$. In this
coordinate system
$$
\{I^i_j\}=\{I^a_b,I^p_q\}, \qquad \{J^i_j\}=\{-I^a_b,I^p_q\}\ .
\eqn\seight
$$
{}From the Nijenhuis conditions satisfied by the complex structures $I,J$ one
gets that $I^a_b=I^a_b(y^a)$ and $I^p_q=I^p_q(z^p)$.  Further analysis of the
conditions \ssix\ in this co-ordinate system reveals that (locally)
$$
\eqalign{ a&=k(y^a)+l(z^p), \qquad e=k(y^a)-l(z^p),
\cr b&=m(y^a)+f(z^p), \qquad b=-m(y^a)+f(z^p)}
\eqn\snine
$$
where $k,l,m,f$ are functions of the indicated co-ordinates which in addition
satisfy
$$
\partial_a k=I^b{}_a\partial_bm, \qquad \partial_pl=I^q{}_p\partial_qf.
\eqn\sten
$$
The second equation of \sseven\ now becomes the two equations
$$
\eqalign{
\partial_x\phi^a&=m\big((\a_+\a_-+\b_+\b_-) g^{ab}+(\a_-\b_+-\a_+\b_-)
I^{ab}\big)\partial_bk
\cr
\partial_x\phi^p&=m\big((\a_+\a_--\b_+\b_-) g^{pq}+(\a_-\b_++\a_+\b_-)
I^{pq}\big)\partial_ql\ ,}
\eqn\seleven
$$
which can be further simplified by introducing complex co-ordinates on the
sigma model manifold with respect to the complex structures $I^a_b, I^p_q$.
There are four topological charges $Q_a, Q_b, Q_c, Q_e$  one for each function
$a,b,c,e$. The values of the topological charges for soliton configurations are
$$
Q_a=a(p_1)-a(p_2), \quad Q_b=b(p_1)-b(p_2), \quad Q_c=c(p_1)-c(p_2), \quad
Q_e=e(p_1)-e(p_2)
\eqn\stwelve
$$
where $p_1,p_2$ are any two vacua between which the soliton solution
interpolates.  The topological charges can also be expressed in terms of the
functions $l,k,m,f$.

In the case that the torsion vanishes ($H=0$), we may choose $I=J$ and $u=da,
v=db, w=dc$ and $n=de$.  One then can show that \btwo\ implies, locally, that
$$
\eqalign{
X_i&=I^j{}_i\partial_j({c-b\over 2})\ , \qquad Y_i=I^j{}_i\partial_j({a+e\over
2})\ ,
\cr
T&=X,\qquad  Z=-Y\ , \qquad a-e=2 (h+{\bar{h}})\ ,\qquad b+c=-2 i (h-
{\bar{h}})\ ,}
\eqn\sthirteen
$$
where $h$ is a holomorphic function. The equations \sfive\ now become
$$
\eqalign{ \partial_t\phi^i-& m
\big[(\a_+\a_-+\b_+\b_-) X^i+(\a_+\b_- -\a_-\b_+) Y^i\big]=0
\cr
\partial_x\phi^i-& m
\big[(\a_+\a_-+\b_+\b_-) g^{ij} \partial_ja + (\a_+\b_- -\a_-\b_+)
g^{ij}\partial_jc
\cr
+ &2(\b_+\b_--i\a_-\b_+)g^{ij}\partial_jh +
2(\b_+\b_-+i\a_-\b_+)g^{ij}\partial_j{\bar{h}}\big]=0\ .}
\eqn\sfourteen
$$
As discussed in the (1,1) case, when the torsion vanishes each Noether charge
$Q$ can be separated into a new Noether charges $Q'$, of the form given in
\sthree, and a topological charge. So this model has two Noether charges $Q'_X$
and $Q'_Y$ and a total of four topological charges $Q_a,Q_b,Q_c$ and $Q_e$.
The values of the latter are given in \stwelve. When $h=0$ two of the four
topological charges are linearly independent and the supersymmetry algebra has
an O(2) automorphism group, as discussed in more detail in [\ustwo]. Thus, the
O(2) invariant models have two Noether and two topological charges and are
hence are possible candidates for self-dual models invariant under the
interchange of Noether and topological charges, as has been suggested [\wo] in
the context of four-dimensional supersymmetric Yang-Mills theory.

A possible realization of (2,2) models for which neither the torsion vanishes
nor all the Killing vectors would be the extension to (2,2) supersymmetry of
the massive group manifold models that we discussed in the previous section.
However, even if one could construct such a model there are good reasons for
believing that the potential could not have any supersymmetric vacua, i.e.
zeros, because for (2,2) models the Witten index is expected to be the sum of
the number of (non-degenerate) supersymmetric vacua, as is the case for the
(2,2) Landau-Ginsburg models. On this basis one would conclude that the Witten
index could not vanish if there were to exist a potential $V$ with any point
for which $V=0$, but the Witten index must vanish when the target space is a
group manifold. In fact, the models discussed in the previous section cannot be
extended to (2,2) supersymmetry for any choice of group manifold as the target
space, despite the fact that there exist massless (2,2) supersymmetric sigma
models on group manifolds for non-zero torsion.


\chapter {(4,4) Models}

To study the equations \eone\ for sigma models with (4,4) supersymmetry we
first consider the case where all vector fields $X,Z_r, Y_s, Z_{sr}$ vanish.
In this case we set
$$
u=da\ ,\quad v=d , \quad w=dc\ , \quad v_{rr}=de_r\ ,\quad v_{sr}=de_{sr}\quad
s\not=r.
\eqn\sfifteen
$$
Provided that $I_rJ_s-J_sI_r=0$, the locally defined functions $b_r, c_s$,
$e_r, e_{sr}$ can be expressed, using \btwo, in terms of $a$ and the complex
structures $I_r$ and $J_s$ as follows:
$$
\eqalign{
\partial_ib_r&=-I_r{}^j{}_i \partial_ja, \quad \partial_ic_s=-J_s{}^j{}_i
\partial_ja
\cr
\partial_ie_r&=\Pi_{rr}{}^j{}_i\partial_ja, \quad \partial_ie_{sr}=
\Pi_{sr}{}^j{}_i\partial_ja \ ,}
\eqn\ssixteen
$$
where $\Pi_{sr}:=I_rJ_s$ are product structures.  The Bogomol'nyi equations are
then
$$
\eqalign{
\partial_t\phi^i&=0,
\cr
\partial_x\phi^i&= m\big[\a_+ \a_- g^{ij} + \a_+ \b_-^s J_s{}^{ij}  +\a_-
\b_+^r
I_r{}^{ij}  + \b_+^r \b_-^s \Pi_{sr}^{ij}\big]\partial_ja.}
\eqn\sseventeen
$$
This model has 16 topological charges each of which is associated with one of
the functions $a, b_r,c_r,e_r, e_{sr}$, but all the Noether charges are zero.
The values of the topological charges for soliton configurations are the
differences of the values of these functions at the vacua between which the
soliton interpolates.  In distinction to the (2,2) models, one cannot choose
coordinates on the target manifold of a generic (4,4) model such that all
complex structures are simultaneously constant; for this reason the Bogomol'nyi
equations \sseventeen\ cannot, in general, easily be further simplified.

We shall now specialize to the case that $H=0$. We may then choose $I_r=J_r$,
set $T_r=Z_{rr}$ and introduce functions $a,b_r,c_s,e_r,e_{sr}$ as in
\sfifteen.  The conditions \btwo\ imply that
$$\eqalign{
X&=T_r,\qquad Y_s=0, \quad Z_r=0, \quad Z_{sr}=0 \ s\not=r,
\cr
c_r&=-b_r,\quad a=0, \quad e_r=0, \quad e_{sr}=0,
\cr
X_i&= I_r{}^j{}_i\partial_jb_r;\quad r=1,2,3.}
\eqn\seighteen
$$
The last of these equations implies that the vector field $X$ is
tri-holomorphic ($\cL_X I_r=0$). The Bogomol'nyi equations can now be written
as
$$
\eqalign{
\partial_t\phi^i&={m\over 2}\eta_{{}_0} X^i,
\cr
\partial_x\phi^i&= {m\over 2}\sum_r\eta_r I_r{}^i{}_j X^j}
\eqn\snineteen
$$
where
$$
\eqalign{
\eta_{{}_0}&=2(\a_+a_-,0,\dots,0,\b_{+1}\b_{-1},\b_{+2}\b_{-2},\b_{+3}\b_{-3})
\cr
\eta_r &=2(0,\dots,0,-\a_+\b_{-r},\b_{+r}\a_-,0, \dots,0) }
\eqn\stwenty
$$
are 16-component vectors and the inner product is defined in eqn. \nice.
Because of equations \restrict\ and the definition of inner product, we find
that
$$
\eta_{{}_0}\cdot \eta_{{}_0}+\sum_r\eta_r\cdot\eta_r=1\ ,
\eqn\stwentyone
$$
and that $\eta_{{}_0}$ and $\eta_r$, $r=1,2,3$, are mutually orthogonal.
Because of the latter property we can change the basis in the 16-dimensional
vector space such that the basis vectors include $\eta_{{}_0}$ and $\eta_r$,
which span a 4-dimensional Euclidean subspace.

As for the (1,1) models, when the torsion vanishes each Noether charge $Q$ can
be separated into a new Noether charge $Q'$ and a topological charge. Thus, the
model under discussion has one Noether charge $Q'_X$ and three topological
charges $Q_r\equiv Q_{b_r}$ associated with the functions $b_r$. These four
charges can be viewed as a 4-vector in the four-dimensional space spanned by
$\eta_{{}_0}$ and $\eta_r$. Clearly, only this four-dimensional subspace of the
original 16-dimensional space is relevant to the analysis to follow.

There is an alternative way to view the topological charges of the theory
using symplectic geometry.  To every complex structure $I_r$ is associated the
K{\"a}hler two-form
$$
{\Omega_r}_{ij}:=-g_{ik} I_r{}^k{}_j\ .
\eqn\stwentytwo
$$
The vector field $X$ is a Hamiltonian vector field of the sympletic form
$\Omega_r$, so
$$
i_X\Omega_r=db_r,
\eqn\stwentythree
$$
{\sl i.e.}  $b_r$ is the Hamiltonian function of $X$ with respect to
$\Omega_r$. By analogy with the Killing potential of a holomorphic Killing
vector field we shall call $b_r$ the hyper-K{\" a}hler Killing potential. The
topological charges are
$$
Q_r=\int\!\! dx\, \phi^* (i_X \Omega_r) = \int\!\! dx\,\partial_x b_r\ .
\eqn\stwentyfour
$$

To construct examples of zero torsion (4,4) massive supersymmetric sigma models
with a 4k-dimensional hyper-K{\" a}hler target space, let the $4k$ coordinates
be $\phi^i=(\varphi^\rho,\phi_r^\rho)$ where $\rho=1,\dots,k$ and $r=1,2,3$.
Using these co-ordinates, we introduce the orthonormal frame
$$
E^{0\rho}= \big(U^{-1/2}\big){}^{\rho\tau}
(d\varphi^\tau+\omega^\tau_{r\lambda}
d\phi^{r\lambda})\qquad
E^{r\rho}= \big(U^{1/2}\big){}^{\rho\tau} d\phi^{r\tau}
\eqn\stwentyfive
$$
and its dual
$$
E_{0\rho}= \big(U^{1/2}\big){}^{\rho\tau} {\partial\over
\partial\varphi^{\tau}}
\qquad
E_{r\rho}= \big(U^{-1/2}\big){}^{\rho\tau}
({\partial\over \partial\phi^{r\tau}}-\omega^\tau_{r\lambda} {\partial\over
\partial \varphi^\lambda})\ ,
\eqn\stwentysix
$$
where $U^{\rho\tau}$ and $\omega^{\rho\tau}$ are the components of $k\times k$
symmetric matrix functions of $\phi^\rho_r$. The metric takes the form [\hr]
$$
\eqalign{
dS^2&\equiv \delta_{ab}\delta_{\rho\mu}E^{a\rho}\otimes E^{b\mu}
\cr
&=U^{\rho\tau} d\phi^\rho\cdot d\phi^\tau + (U^{-1})^{\rho\tau}(d\varphi^\rho
+\omega^{\rho\lambda}\cdot d\phi^\lambda) (d\varphi^\tau +\omega^{\tau\mu}\cdot
d\phi^\mu)  \ ,}
\eqn\stwentyseven
$$
where $\big(a=(0,r),  b=(0,s)\big)$. The hyper-K{\"a}hler condition requires
that $U$ and $\omega$ satisfy
$$
{\partial\over\partial\phi^\rho_r}\omega_s^{\lambda\tau}-
{\partial\over\partial\phi^\lambda_s}\omega_r^{\rho\tau}=
\varepsilon_{rst} {\partial\over\partial\phi^\rho_t}U^{\lambda\tau}\qquad
{\partial\over\partial\phi^\tau_r}U^{\rho
\lambda}={\partial\over\partial\phi^\rho_r}U^{\tau \lambda}\ .
\eqn\stwentyeight
$$

The three closed K{\"a}hler 2-forms are
$$
\eqalign{
\Omega_r&\equiv D^{(-)}(T_{0r})_{ab} \delta_{\rho\tau} E^{a\rho}\wedge
E^{a\tau} \cr
&=2 d\varphi^\rho\wedge d\phi_r^\rho +( 2 \omega_s^{\rho\tau}\delta_{tr}
-U^{\rho\tau}\varepsilon_{rst})d\phi_s^\rho\wedge d\phi_t^\tau\; \ ,}
\eqn\stwentynine
$$
where
$$
D^{(\pm)}(T_{ab})= D(T_{ab})\pm {1\over2} \epsilon_{ab}{}^{cd}D(T_{cd})\ ,
\eqn\sthirty
$$
and
$$
D(T_{ab}){}^c{}_d= \delta_{ad} \delta_b{}^c-\delta_{bd} \delta_a{}^c\ ,
\eqn\sthirtyone
$$
{\sl i.e.} $D^{(-)}(T_{0r})$ is the anti-self dual part of the fundamental
representation of $\cL \big(SO(4)\big)$ restricted to one of the two  $\cL
\big(SO(3)\big)$ subalgebras of $\cL \big(SO(4)\big)$. Observe also that
$$
D^{(\pm)}(T_{0r}) D^{(\pm)}(T_{0s})=-\delta_{rs}\pm \epsilon_{rst}
D^{(\pm)}(T_{0t})\ .
\eqn\sthirtytwo
$$
The associated complex structures are
$$
I_r=-D^{(-)}(T_{0r}){}^a{}_b {}\delta^\rho{}_\tau{} E^{b\tau}\otimes
E_{a\rho} \ .
\eqn\sthirtytwoa
$$
A basis for the $k$ triholomorphic Killing vectors is
$\big\{{\partial\over\partial\varphi^\rho}\big\}$. If we choose the particular
linear combination
$$
X=c^\rho{\partial\over\partial\varphi^\rho}
\eqn\dtwentythree
$$
for the Killing vector of the massive sigma-model \ssaction\ then it is easily
seen that the hyper-Killing potential is just
$$
b_r=c^\rho\phi^\rho_r\ .
\eqn\dtwentyfour
$$
The Bogomol'yni equations \snineteen\ now reduce to
$$
\eqalign{
\partial_t\varphi^\rho&={m\over2}\eta_{{}_0} c^\rho\qquad \qquad
\partial_t\phi^\rho_r =0
\cr
\partial_x\varphi{}^\rho&=-{m\over2} c_\sigma\big(U^{-1}\big)^{\sigma\tau}
\eta^r
\omega_{r\tau}^\rho  \qquad
\partial_x\phi_r^\rho  ={m\over2} \big(U^{-1}\big)^{\rho\tau}c_\tau
\eta_r \ .}
\eqn\dtwentyfive
$$
Using the equation for $\phi^\rho_r$ the equation for $\varphi$ can be
rewritten as
$$
\partial_x\varphi{}^\rho=- \partial_x\phi^{r\tau} \omega^\rho_{r\tau}\ .
\eqn\dtwentysix
$$
The first two Bogomol'yni equations of \dtwentyfive \ can be solved easily as
follows:
$$
\varphi^\rho(x,t)={m\over2}\eta_{{}_0} c^\rho t+\varphi^\rho(x)
\qquad
\phi^\rho_r(x,t) =\phi^\rho_r(x)\ .
\eqn\dtwentyseven
$$
Substituting the solutions \dtwentysix\ into the other Bogomol'yni equations,
we get the following independent ordinary differential equations:
$$
{d\over dx}\phi_r^\rho(x)  ={m\over2} \big(U^{-1}\big)^{\rho\tau}c_\tau \eta_r
\qquad
{d\over dx}\varphi^\rho(x)=- {d\over dx}\phi^{r\tau} \omega^\rho_{r\tau}\ .
\eqn\dtwentyeight
$$
The only non-trivial differential equation to be integrated is that for
$\phi_r^\rho$. This is because if the solution for $\phi_r^\rho$ is known then
the solution of the equation for $\varphi{}^\rho(x)$ can be found from
$$
\varphi^\rho(x)=- \int \! dx\! {d\over dx}\phi^{r\tau} \omega^\rho_{r\tau}+
\varphi^\rho_0
\eqn\dtwentynine
$$
where $\varphi^\rho_0$ is a constant, {\sl i.e} $\varphi^\rho(x)$ is the
holonomy of the connection $\omega^\rho_{r\tau}$.  Note that
$\omega^\rho_{r\tau}$ depends only on $\phi_r^\rho$.

When the target space is four-dimensional the metrics of
\stwentyseven\ are the Gibbons-Hawking multi centre 4-metrics [\GH]
$$
dS^2 = Ud\phi\cdot d\phi + U^{-1}\big( d\varphi + \omega\cdot d\phi\big)^2
\eqn\dthirty
$$
where $\varphi$ is an angular variable, $\omega$ and $\phi$ are now considered
as vectors in Euclidean three-space, and ${\rm curl}\, \omega= {\rm grad}\, U$.
An explicit soliton solution was given in [\Qkinks] for the special case of the
two-centre metric
$$
U= 2\mu\Big[ {1\over |\phi -\phi_0|} +  {1\over |\phi +\phi_0|}\Big]
\eqn\dthirty
$$
($\mu>0$, $0\leq \varphi\leq 8\pi \mu$), and triholomorphic Killing vector
$X={\partial\over\partial\varphi}$ (so that $V={m^2 \over 4}U^{-1}$). We
conclude by presenting some further aspects of this solution. A solution of
${\rm curl}\, \omega= {\rm grad}\, U$ for $U$ given by \dthirty\ is
$$
\eqalign{
\omega= &\mu \big(\phi \times \phi_0\big)\Bigg\{  {|\phi_0|-|\phi-\phi_0|\over
|\phi|^2|\phi_0||\phi-\phi_0|}- {|\phi_0|-|\phi+\phi_0|\over
|\phi|^2|\phi_0||\phi+\phi_0|}
\cr
+& {\phi\cdot\phi_0\over |\phi|^2} \Big[ {1\over
\big(|\phi||\phi-\phi_0|+\phi\cdot (\phi-\phi_0)\big)|\phi-\phi_0|}- {1\over
(|\phi||\phi-\phi_0|
-\phi\cdot\phi_0)|\phi_0|}
\cr
+& {1\over
\big(|\phi||\phi+\phi_0|+\phi\cdot (\phi+\phi_0)\big)|\phi+\phi_0|}-
{1\over(|\phi||\phi+\phi_0|
+\phi\cdot\phi_0)|\phi_0|}\Big]\Bigg\}\ . }
\eqn\dthirtyone
$$
Note that this $\omega$ is orthogonal to $\phi_0$. Since $Q=2\phi_0$ for this
example, and $\eta$ is parallel to $Q$ for a solution saturating the
Bogomol'nyi bound, we conclude that $\eta\cdot\omega =0$ and \dtwentynine\
reduces to
$\varphi(x)=\varphi_0$. The solution for $\phi$ is straightforward and is
$$
\phi^r = \phi_0^r \tanh \Big({ m|\eta|\over 8\mu}(x-x_0)\Big)\ .
\eqn\dthirtyone
$$
This is the `Q-kink' solution found in [\Qkinks]. Note that $\omega$ is
determined only up to the addition of the gradient of a scalar function. For a
generic solution of ${\rm curl}\, \omega= {\rm grad}\, U$, the 3-vector
$\omega$ would not be orthogonal to $\eta$, and $\varphi$ would not be
constant. However, it is possible to redefine $\varphi$ to compensate for a
gradient term in $\omega$ and this means that, at least for the two-centre
metric, we may always arrange for $\varphi$ to be constant for the soliton
solution.

The moduli space of this soliton solution is $\RN \times S^1$ with co-ordinates
$(x_0, \varphi_0)$. The metric on this space may be computed from \efour\ and
is
$$
ds^2 = {1\over2}|\eta|^2 \Big( {m|\phi_0|\over |\eta|} dx_0^2 + {4|\phi_0|\over
m|\eta|^3}d\varphi_0^2\Big)\ .
\eqn\dthirtytwo
$$
Defining the new variable $y$ by
$$
\varphi_0 = {1\over2}m\, y \ ,
\eqn\dthirtythree
$$
and introducing the notation
$$
\gamma \equiv |\eta|^{-1} ={1\over \sqrt {1-\eta_{{}_0}^2}}\ , \qquad
\qquad M=m|\phi_0|\ ,
\eqn\dthirtyfour
$$
we find that
$$
\gamma^2 \big({ds\over dt}\big)^2 = {1\over2}\Big[
\gamma M
\dot x_0^2 +
\gamma^3 M\dot y^2\Big]\ .
\eqn\dthirtyfive
$$
If we interpret $\eta_{{}_0}$ as the velocity of the particle in the $y$
direction, then $\gamma$ is the time dilation factor associated with this
motion, so $Q_X= \gamma\eta_{{}_0}|Q|$ is the momentum in this direction and
the total energy $E= \gamma M$ is the inertial mass for acceleration in
orthogonal directions, i.e. in the $x_0$ direction. In this interpretation the
inertial mass in the $y$ direction would be expected to be $\gamma^3 M$ and
this is precisely the coefficient of the $\dot y^2$ term in \dthirtyfive. Thus,
the right hand side of \dthirtyfive\ may be interpreted as the bosonic part of
the soliton's full effective worldline action which will have an $N=4$
worldline supersymmetry. Note that the left hand side of \dthirtyfive\ can be
rewritten in
terms of the soliton's proper time $\tau$, defined by $d\tau= \gamma^{-1}dt$,
as
$(ds/d\tau)^2$.

This `Kaluza-Klein' interpretation can be understood as a consequence of the
fact that this massive sigma model model can be obtained by (non-trivial)
dimensional reduction of a massless model in three dimensions, as explained in
[\Qkinks]. In contrast, because of the torsion, the massive (1,1) model of
section 4 cannot be so obtained and the effective Lagrangian  \setwenty\ is not
of the same form. Indeed, while it is tempting to identify $\kappa$ in
\seeighteen\ as the velocity in the extra dimension this identification is not
consistent with the interpretation of $Q_X$ as the momentum in this direction
since $Q_X$ does not vanish for $\kappa=0$. Also, whereas the massive (4,4)
sigma model action does not depend on the `velocity' $\eta_{{}_0}$, the massive
(1,1) sigma model action does depend on $\kappa$, which is therefore fixed by
the model.


\chapter {Summary}

In previous work we have established the conditions required for (p,q)
supersymmetry of the most general `massive' supersymmetric sigma-model, by
which we mean the inclusion into a massless model of a potential term $V(\phi)$
for the sigma-model scalar fields $\phi$, and the concomitant Yukawa couplings.
The question then arises of whether massive models can be found satisfying
these conditions. In the absence of torsion the conditions imposed by (p,q)
supersymmetry are simpler than those for non-vanishing torsion, and various
supersymmetric massive models with zero torsion had been constructed
previously. In this paper we have constructed a new class of (1,1)
supersymmetric massive sigma models with a group manifold as the target space,
and non-zero torsion; a subclass can be considered as a massive extension of
the well-known supersymmetric WZW models. It is possible to show the existence
of massive (2,2) models with torsion but it seems that there are no massive
(2,2) supersymmetric models with a group manifold as the target space.  We have
also constructed a class of (4,4) models with 4k-dimensional target spaces, but
without torsion,
that generalize previously constructed models based on the Gibbons-Hawking
4-metrics.

In many cases the potential $V$ has multiple isolated zeros, which are
necessarily (supersymmetric) ground states. A general feature of models with
such potentials is the existence of soliton solutions interpolating between
these distinct vacua. Using the (p,q) supersymmetry algebra of the
supersymmetry charges obtained in our previous work, we have shown that the
solitons saturate a Bogomol'nyi bound on the energy, in terms of the
topological and Noether charges which occur in the supersymmetry algebra as
central charges. In simple cases for which the potential $V$ has just two
isolated zeros, exact solutions can be found and we presented a new soliton
solution of a particular
massive (1,1) sigma model with the group $SU(2)$ as its target space.

For the explicit soliton solutions considered we have also computed the mass
and charges, determined the metric on the moduli space and the (bosonic part of
the) associated effective Lagrangian for one soliton. Models for which the
potential $V$ has multiple vacua will generically allow multi-soliton
solutions, although these solutions will generally be time-dependent and need
not be solutions of the Bogomol'nyi equations (e.g. the supersymmetric
sine-Gordon model). We leave to the future the determination of which of the
new models introduced here have this feature.

\vfill\eject


\noindent{\bf Acknowledgements:} G.P. was supported by a Royal Society
University Research
Fellowship.

\refout

\bye